\documentclass[10pt,a4paper,onecolumn]{article} 

\usepackage[pdftex]{graphicx}
\usepackage{amssymb}
\usepackage{amsmath}
\usepackage[square,comma,numbers,sort&compress]{natbib}
\usepackage{url}
\usepackage[font={small}, labelfont=bf]{caption} 

\newcommand{\f}{\operatorname}

\usepackage{hyperref}

\usepackage{environ}         
\usepackage{etoolbox}        
\usepackage{graphicx}        
\newlength{\myl}
\let\origequation=\equation
\let\origendequation=\endequation
\RenewEnviron{equation}{
  \settowidth{\myl}{$\BODY$}  
  \origequation
  \ifdimcomp{\the\linewidth}{>}{\the\myl}
  {\ensuremath{\BODY}}                             
  {\resizebox{\linewidth}{!}{\ensuremath{\BODY}}}  
  \origendequation
}

\title{Power laws in the Roman Empire: a survival analysis}

\author{
Pedro L. Ramos$^1$, Luciano da F. Costa$^2$, Francisco Louzada$^1$, Francisco A. Rodrigues$^1$ \\
\normalsize{$^{1}$Institute of Mathematical Science and Computing, University of S\~ao Paulo, S\~ao Carlos, Brazil}\\
\normalsize{$^{2}$S\~ao Carlos Institute of Physics, University of S\~ao Paulo, S\~ao Carlos, Brazil}
}

\begin{document}

\maketitle

\begin{abstract}
The Roman Empire shaped Western civilization, and many Roman principles are embodied in modern institutions. Although its political institutions proved both resilient and adaptable, allowing it to incorporate diverse populations, the Empire suffered from many internal conflicts. Indeed, most emperors died violently, from assassination, suicide, or in battle. These internal conflicts produced patterns in the length of time that can be identified by statistical analysis. In this paper, we study the underlying patterns associated with the reign of the Roman emperors by using statistical tools of survival data analysis. We consider all the 175 Roman emperors and propose a new power-law model with change points to predict the time-to-violent-death of the Roman emperors.  This model encompasses data in the presence of censoring and long-term survivors, providing more accurate predictions than previous models. Our results show that power-law distributions can also occur in survival data, as verified in other data types from natural and artificial systems, reinforcing the ubiquity of power law distributions. The generality of our approach paves the way to further related investigations not only in other ancient civilizations but also in applications in engineering and medicine.
\end{abstract}

\section{Introduction} \label{sec:introduction}

As complementation to continuing studies by historians, anthropologists, and social scientists, ancient civilizations can be analyzed quantitatively, considering mathematical and computational modelling~\cite{richardson1960statistics, bologna2008simple}. Ancient civilizations are examples of a complex system composed of agents that interact, collaborate, and compete for power and resources. One example of such a system is the Roman Empire, which started with Augustus (d. 14 CE) and ended with Romulus Augustulus, with the Germanic invasion from the North in 476. After this, the Eastern Roman Empire, with Constantinople as its capital, continued to exist until 1453~\cite{Retief06}. The Roman Empire influenced western civilization and contributed to many aspects of our culture~\cite{Rostovtzeff26}. Although the first two centuries of the Empire were characterized by stability and prosperity, known as the Pax Romana, the Roman Empire underwent several crises, including many violent deaths of Roman emperors. A recent study suggested that among the 69 rulers, 43 emperors suffered a violent death~\cite{saleh2019statistical}, including homicide, suicide, or death in combats with a foreign enemy of Rome. These internal conflicts produced patterns in the length of time that can be identified by statistical analysis, as we show in this paper. 

We propose a power-law model to describe the time-to-violent-death, which measures the time before a Roman emperor suffers a violent death since the begging of its reign. This variable is interesting from the historical point of view as it can suggest some patterns that governed the dynamics of the Roman Empire. For example, if deaths occur independently, without any defined rule, we should observe an exponential distribution of this survival time, where the hazard function is constant. Otherwise, some other distribution should be more suitable to model the time-to-violent-death. Recently, Saleh~\cite{saleh2019statistical} analyzed the time-to-violent-death of the 69 emperors of the unified Roman Empire, from Augustus (d. 14 CE) to Theodosius (d. 395 CE). The author considered statistical tools of survival data analysis to model their time-to-violent-death and verified that a mixture of two Weibull distributions well captures this variable. However, as we show here, that model does not seem to be easily extended to the long-term
perspective observed for some Roman Emperors who did not face a violent death. Moreover, the dataset analyzed covers only part of the Roman emperors, leaving out, for instance, the Byzantine Empire. To address these limitations, we show that a power-law distribution with change points can better model the time-to-violent-deaths. This power-law behaviour in time
distribution has been observed in many complex systems, including bacterial persistence~\cite{csimcsek19}, earthquakes~\cite{bak2002unified, Mega03, kagan2010earthquake}, in solar radiophysics~\cite{Gary04}, stock price fluctuations \cite{gabaix2009power}, and tree-limb branching \cite{bentley2013empirical}. The cause of this distribution is related to the presence of extreme events, such as large earthquakes. In the case of the Roman Empire, we observe that an extreme event is an emperor survive for many years since the begging of his reign. Our model shows that most of Rome's emperors suffered violent deaths during the initial years of their reign.

We study the distribution of time-to-violent-death based on survival analysis, which refers to a set of statistical approaches used to investigate the time it takes for an event of interest to occur~\cite{miller2011survival}. The term survival analysis originates from clinical research, where predicting the time to death, i.e., survival, is often the primary goal. The maximum likelihood estimator (MLE) is used to estimate the best distribution for all the considered emperors, considering data in the presence of censoring and long-term survivors. Indeed, the inclusion of long-term survival enables a more accurate prediction regarding the emperors who did not have a violent death than in the study by Saleh~\cite{saleh2019statistical}. By considering the dataset by Saleh~\cite{saleh2019statistical},  we also verify that the risk after assuming the throne is very high for a Roman emperor, but this risk systematically decreases until 13 years of rule rapidly increases after this change point. This change point indicates a change in the Roman Empire's tendency and dynamics after a given time of reign. Further, we also consider the Byzantine Empire's time-to-violent-death that, combined with the initial data, comprehends 175 emperors. In this case, we observe a more complex behavior with additional change points and, therefore, we generalize the power-law model to include multiple change points. 

Our model is mainly general and can be used in any system in which the reliability function follows a power law, and we have censored data. Indeed, our work opens the possibility to study empires and dynasties by using statistical analysis tools, contributing to the understanding of social and historical sciences more quantitatively. 

This paper is organized as follows. Initially, we consider the data by Saleh~\cite{saleh2019statistical} and show that the power-law model is more suitable to model the distribution of time-to-violent-death than previous models. We identify a change point that occurs at 13 years of reign. Next, we consider a data set of 175 Roman emperors, including also the Byzantine Empire. In this case, we introduce a generalized power-law distribution with $k-1$ change points and show that this distribution accurately describes the data observed.  Finally, we analyze some possible causes of the short death of Roman's emperor and verify that the birthright is a statistically significant variable, whereas other features, such as the birth province and dynasty, do not influence the time-to-violent-death. These findings suggest that emperors who inherited the reign tend to have a more peaceful reigning and increased probability of natural death. At the end of the paper, we present the mathematical formulation and simulation study related to our statistical model.

\section{Survival analysis}

The power-law distribution can be represented by its complementary cumulative density  function (also refereed by survival function) given by
\begin{equation*}
S(x)=\left( \frac{x}{x_{\min}}\right)^{1-\alpha},
\end{equation*}
where $\alpha>1$ and $x_{\min}>0$ is the minimum value at which power-law behavior holds. The inferential methods to estimate $\alpha$ have been discussed by many authors \cite{clauset2009power,goldstein2004problems,voitalov2019scale}.  The most common approach to this end is the maximum likelihood estimator, which in the continuous case has a simple closed-form expression as well as desired properties, such as consistency and asymptotically efficiency. 

\begin{figure}[!t]
	\centering
	\includegraphics[width=0.7\textwidth]{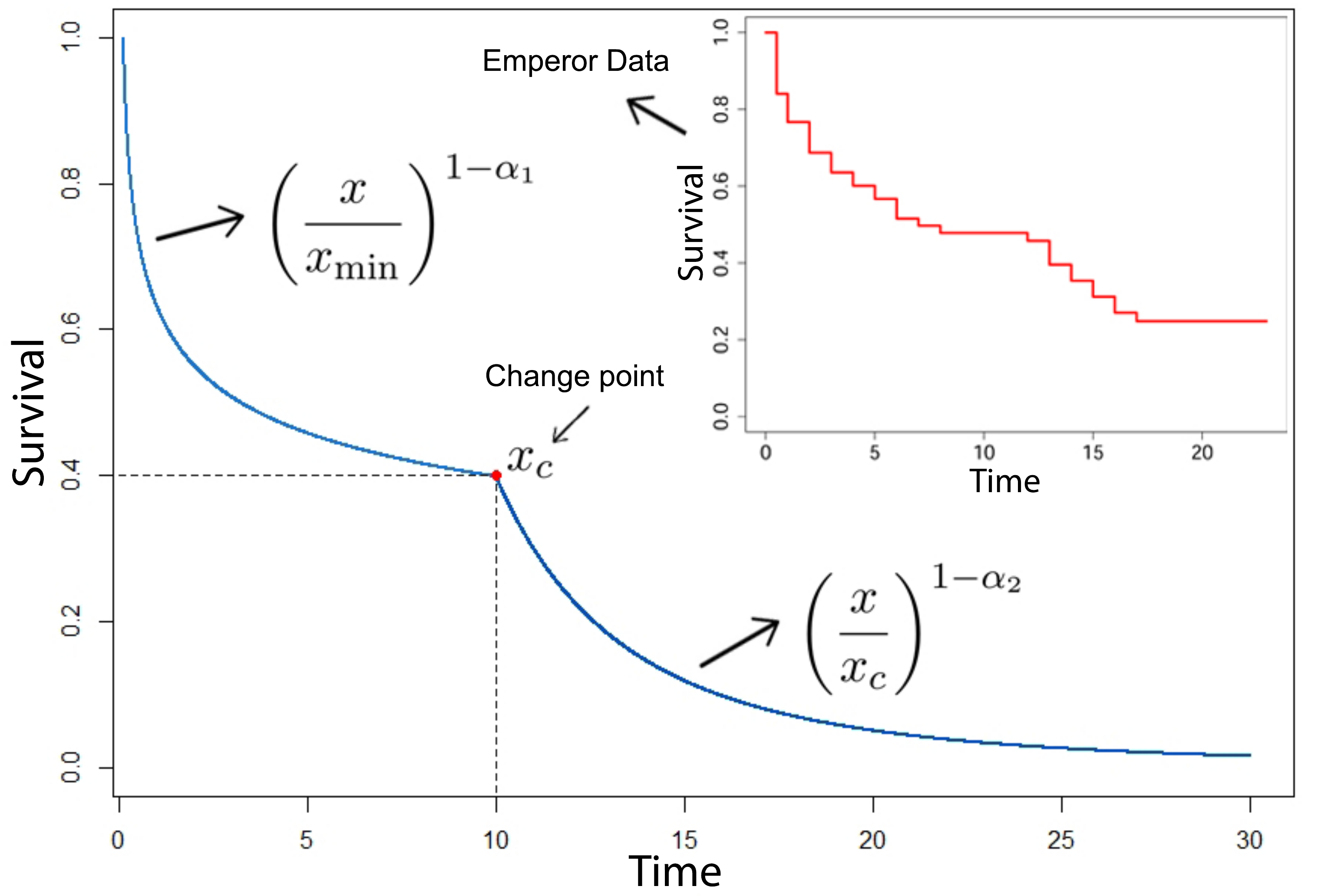}
	\caption{Illustration of the survival function modelled by a power-law distribution with change point $x_c$. The Kaplan-Meier estimator is shown inside the figure.}\label{fsimulation1}
\end{figure}

To construct our model, let us consider Figure \ref{fsimulation1}. The top panel presents the Kaplan-Meier non-parametric estimator for the time-to-violent-death of the 69 rulers of the Roman Empire~\cite{saleh2019statistical}. The Kaplan-Meier estimator shows the probability of an event at a certain time interval. It can be noticed that the survival time follows a power-law distribution, and after a change point, the behavior follows a power-law distribution with a different scaling exponent. In Figure \ref{fsimulation1}, we consider a power-law distribution with $\alpha_1$ parameter and after a critical value $x_c$, the behavior change to a power-law distribution  with coefficient $\alpha_2$.

In a general form, the power law model behavior in Figure~\ref{fsimulation1}  can be represented by the following expression:
\begin{equation}\label{eqpc1}
S(x)=\left( \frac{x}{x_{\min}}\right)^{1-\alpha_1}\mathbb{I}[x_{\min},x_{c}]+C\left( \frac{x}{x_{c}}\right)^{1-\alpha_2}\mathbb{I}(x_{c},\infty),
\end{equation}
where $\mathbb{I}(\cdot)$ is an identity function that return 1 if the value is inside the interval and 0 otherwise, $C=\left( \frac{x_c}{x_{\min}}\right)^{1-\alpha_1}$ is the normalized constant that make the model to be a complementary cumulative function and $x_c>x_{\min}$. From (\ref{eqpc1}) the probability distribution of the power law distribution with change point is given by
\begin{equation}\label{eqpc2}
f(x)=\frac{\alpha_1-1}{x_{\min}}\left( \frac{x}{x_{\min}}\right)^{-\alpha_1}\mathbb{I}[x_{\min},x_{c}]+C\frac{\alpha_2-1}{x_{c}}\left( \frac{x}{x_{c}}\right)^{-\alpha_2}\mathbb{I}(x_{c},\infty).
\end{equation}

An important aspect of the proposed model is that it can be seen as a generalization of the power-law distribution in the thermodynamic limit of $x_c$, i.e.,
$$\lim_{x_{c}\rightarrow\infty}S(x)=\left( \frac{x}{x_{\min}}\right)^{1-\alpha_1},$$
which leads to the standard power-law distribution.  

In many cases, it is useful to present other mathematical functions related to the proposed model, such as the mean, variance, and kurtosis, to name a few. These functions can be obtained from the r-$th$ moment that is derived in section~\ref{Sec:methods}. Another aspect that needs to be discussed is the inferential procedure used to obtain the parameter estimates. Although the maximum likelihood estimator has already been discussed for the power-law distribution (see Clauset et al. \cite{clauset2009power} and the references therein), this method needs to be generalized to our model. The closed-form estimators for the parameters and the asymptotic confidence intervals obtained from the maximum likelihood estimators are discussed in detail in section~\ref{Sec:methods}.

The estimators presented in section~\ref{Sec:methods} can be used satisfactorily in most scenarios. However, these estimators may return biased estimates in the presence of events with incomplete information. For instance, Augustus, known as the founder of the Roman Principate, died at the age of 75, probably due to natural causes. Since he never experienced the event of interest, i.e., ``a violent death'', we only have partial knowledge about his reign, which lasted 40 years. By considering this information as complete, we may obtain biased estimates, as performed in~\cite{saleh2019statistical}. The same problem will occur if we discard this information. On the other hand, statistical analysis has a way to take into account such information during the parameter estimation, i.e., the MLEs in the presence of random censoring, as discussed in detail in section \ref{rancesnmle}. We use this estimate in the data analysis presented in the next section.

\section{Results}

We divide our research into two parts. Initially, we analyze the data by Saleh \cite{saleh2019statistical}, which considers the Western Roman Empire, from Augustus (d. 14 CE) to Theodosius (d. 395 CE). In this analysis, we compare our model with the previous model discussed in~\cite{saleh2019statistical}. Our study includes the Eastern Roman Empire, which ended with Constantine XI Palaiologos (d. 1453 CE). While the first data set considers only 69 emperors, the second one is complete, covering 175 Roman emperors. 

\subsection{Western Roman Empire}

Initially, we consider the data organized by Saleh \cite{saleh2019statistical}, which was obtained from De Imperatoribus Romanis, an online encyclopedia of Roman emperors available as Supplemental Material in the cited paper. Here, we consider the same data set in order to compare the two models. Saleh \cite{saleh2019statistical} has considered a mixture of Weibull distributions to describe the lifetime, given by
\begin{equation}\label{Eq:Saleh}
\hat{S}_{MW}(x)=0.876\cdot\exp\left[-\left(\frac{x}{12.835}\right)^{0.618} \right]+0.124\cdot\exp\left[-\left(\frac{x}{14.833}\right)^{13.387} \right].
\end{equation}

An important characteristic that should be taken into account when analyzing data from the Roman Empire is the existence of emperors who did not have a violent death, i.e., a portion of the emperors was not susceptible to the event of interest. From the model (equation~\ref{Eq:Saleh}), we have that $S(t)\rightarrow0$ as the time increases. Hence, the model assumes that all the emperors will experience a violent death at some time. This assumption is not plausible, as many emperors died due to natural causes and did not have a violent death. Such characteristics can be incorporated into our power-law model by considering a long-term survival model. In this case, $S(t)\rightarrow\pi$ as the time increase, where $\pi\geq 0$ is the proportion of emperors that did not experience a violent death. The details related to the modifications in the power-law distribution with a change point to include long-term survival can be seen in section~\ref{Sec:methods}.

Fitting the power-law distribution to the data, the MLEs are obtained and return the estimates $\hat\alpha_1=1.41$, $\hat\alpha_2=5.12$, and $\pi=0.22$ when fixing $x_{\min}=0.5$ and $x_c=13$. Although these results returned good shapes for the survival function we observe that there is space for improvement. Considering a simple grid search around the obtained values we achieve the following estimates $\hat\alpha_1=1.382$, $\hat\alpha_2=8.5$, and $\pi=0.245$ which provided better curves when compared with the empirical survival function. Hence, the adjusted model is given by:
\begin{equation*}
\hat{S}_{PL}(x) = 0.248+0.752\cdot\left( \frac{x}{0.5}\right)^{-0.382}\mathbb{I}[0.5,13]+  0.522\cdot\left( \frac{x}{13}\right)^{-7.5}\mathbb{I}(13,\infty).
\end{equation*}

\begin{figure}[!t]
	\centering
	\includegraphics[width=0.8\textwidth]{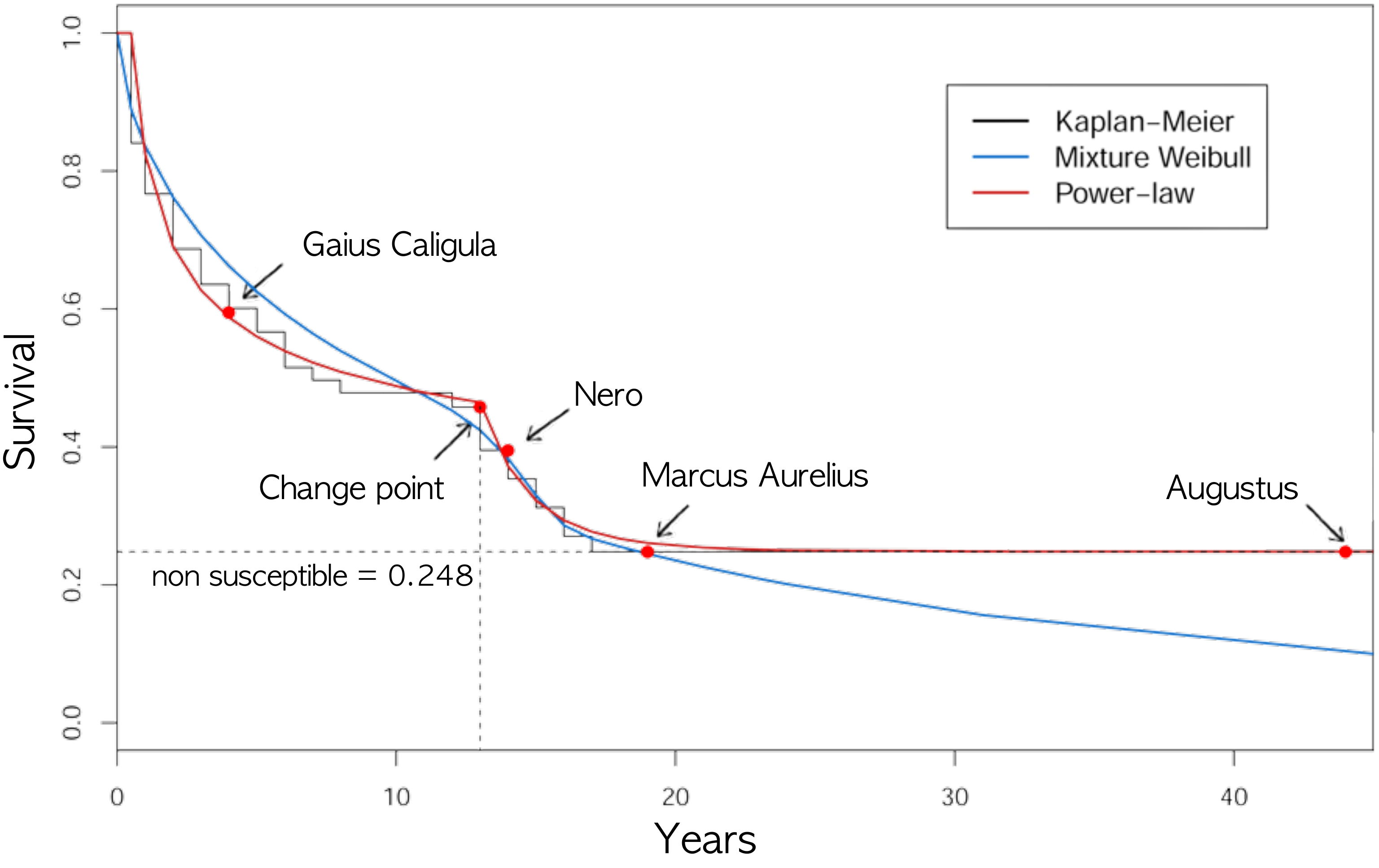}
	\caption{Power-law survivor (reliability) function of Roman emperors, the adjusted mixture Weibull and the nonparametric Kaplan-Meier estimates. We consider $x_{\min}=0.5$, $\alpha_1=1.382$, $x_{c}=13$ and $\alpha_2=8.5$ in the power-law model shown in equation (\ref{eqpc1}).}
\label{fsimulation2}
\end{figure}

Due to discontinuity in the PDF, the standard maximum likelihood of $x_c$ cannot be achieved. This problem can be easily overcome by considering a grid search for the parameter where the Kolmogorov-Smirnov test is used to estimate $x_c$. Figure \ref{fsimulation2} presents the fit of our proposed model compared with the mixture Weibull and the non-parametric Kaplan-Meier estimator.

	\begin{figure}[!b]
	\centering
	\includegraphics[width=0.8\textwidth]{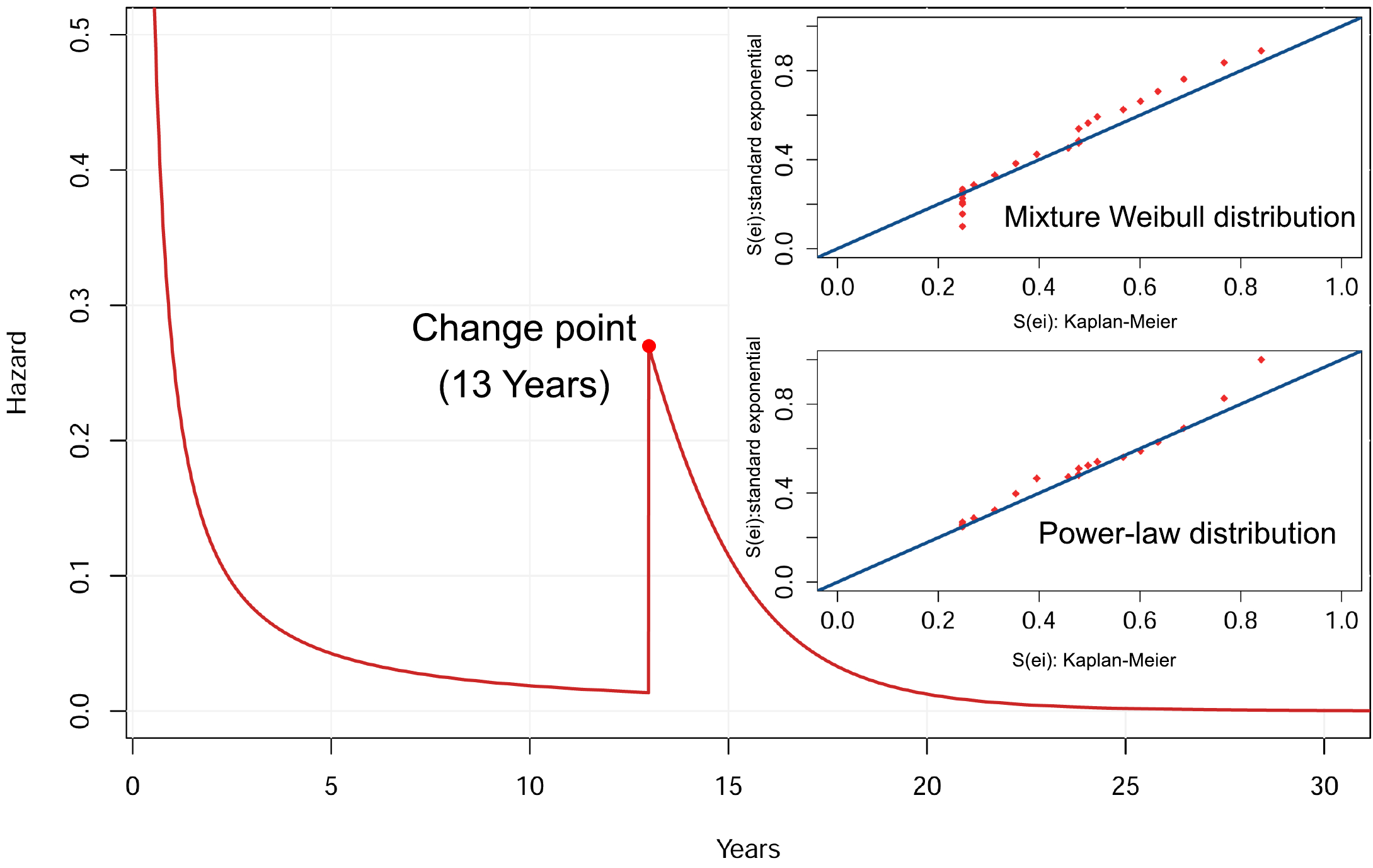}
	\caption{Failure rate from power-law with phase transition distribution of Roman emperors. Cox-snell residuals analysis for the Mixture Weibull and power-law distribution.  }\label{ajustkm}
	\end{figure}

Our results indicate that the change point occurs at 13 years of reign. An interesting fact is that, although Alexander the Great (d. 323 BC) has not participated in the unified Roman Empire and was not included in our analysis, his reign had  13 years. From the adjusted results, we observe that Gaius Caligula (d. 41) had violent death before the change point, while Nero (d. 68 AC) experienced this type of death after the change point. Although we have observed about $37\%$ of censored information, our model estimate that nearly $25\%$ ($\pi=0.248$) of the emperors would not die due to the violent death. For example, Marcus Aurelius (d. 160 AC), known as the last of the five good emperors, lived in an age of relative peace and died due to natural causes. As mentioned earlier, Augustus (d. 14) also reigned for 40 years and did not have a violent death. While our model can capture this effect in the survival function, the mixture Weibull distribution cannot return very far estimates when compared with the non-parametric survival, which is undesirable.

Figure \ref{ajustkm} presents the failure rate of the adjusted model using $h_{\rm pop}(t) =f_{\rm pop}(x)S_{\rm pop}(x)^{-1}$ (given in  equations (\ref{fcp1}) and (\ref{fcp2})) which shows a decreasing failure rate in the first years up to 13 years. During the change point, there is a rapid increase in the hazard of being death (violent) and then decrease again over time. These results differ from the ones obtained from the failure rate of the mixture Weibull distribution. In the latter, the increase in the hazard starts at nearly 11 years, reaches a peak at 15 years, and then decreases. Hence, our proposed approach returns a better understanding of the stochastic process that describes the time-to-violent-death of the emperors of the unified Roman Empire.

\subsection{Western and Eastern Roman Empire}

Although the analysis by Saleh \cite{saleh2019statistical} considered the dynasties that started from Augustus (d. 14 CE) and ended with Theodosius (d. 395 CE) when the Western Roman Empire began
to disintegrate, here, we extend his study considering the period of the  Byzantine Empire (Eastern Roman Empire) that ended with Constantine XI Palaiologos (d. 1453 CE). The additional data included 106 emperor, where the observed censorship was $56.57\%$, which is higher than the observed in the first analyzed period, and the long-term survival was estimated in $\pi=0.404$, i.e., $40.40\%$ of the Roman emperors probably would not face a violent death. This finding implies that during the Byzantine Empire, the emperors were subjected to a less violent death. This change in the hazard behavior may be explained due to different causes; for instance, the eastern half of the Roman Empire showed to be less susceptible to external attacks due to its geography. Additionally, the Empire did not have many common boundaries with Europe. Although many other complex causes are also responsible for providing such stability, those facts combined with the internal political cohesion and more robust administrative center allowed the Byzantine Empire to survive for many centuries. 

Here, the survival function related to the time-to-violent-death for all emperors has more complex behavior, therefore, we have generalized the power-law distribution with $k-1$ change points that can be represented by the following expression:
\begin{equation}\label{eqpc1l}
S(x)=\sum_{i=1}^{k}\left( \frac{x}{x_{(i-1)}^{*}}\right)^{1-\alpha_i}C_{i-1}\mathbb{I}[x_{(i-1)}^{*},x_{(i)}^{*}] \quad \mbox{ and } \quad C_{j}=\prod_{j=1}^{i}\left( \frac{x_{(j)}^{*}}{x_{(j-1)}^{*}}\right)^{1-\alpha_j}
\end{equation}
where $\mathbb{I}(\cdot)$ is an identity function that return 1 if the value is inside the interval and 0 otherwise,  $C_{0}=1$, $x_{(k)}^{*}=\infty$, $\alpha_i>0$ and $x_{(i)}^{*}>x_{(i-1)}^{*}, \forall i=1,\ldots,k$. From (\ref{eqpc1l}) the probability distribution of the piecewise power-law model distribution is given by
\begin{equation}
f(x)=\sum_{i=1}^{k}\frac{\alpha_i-1}{x_{(i-1)}^{*}}\left( \frac{x}{x_{(i-1)}^{*}}\right)^{-\alpha_i}C_{i-1}\mathbb{I}[x_{(i-1)}^{*},x_{(i)}^{*}].
\end{equation}

\begin{figure}[!t]
	\centering
	\includegraphics[width=0.9\textwidth]{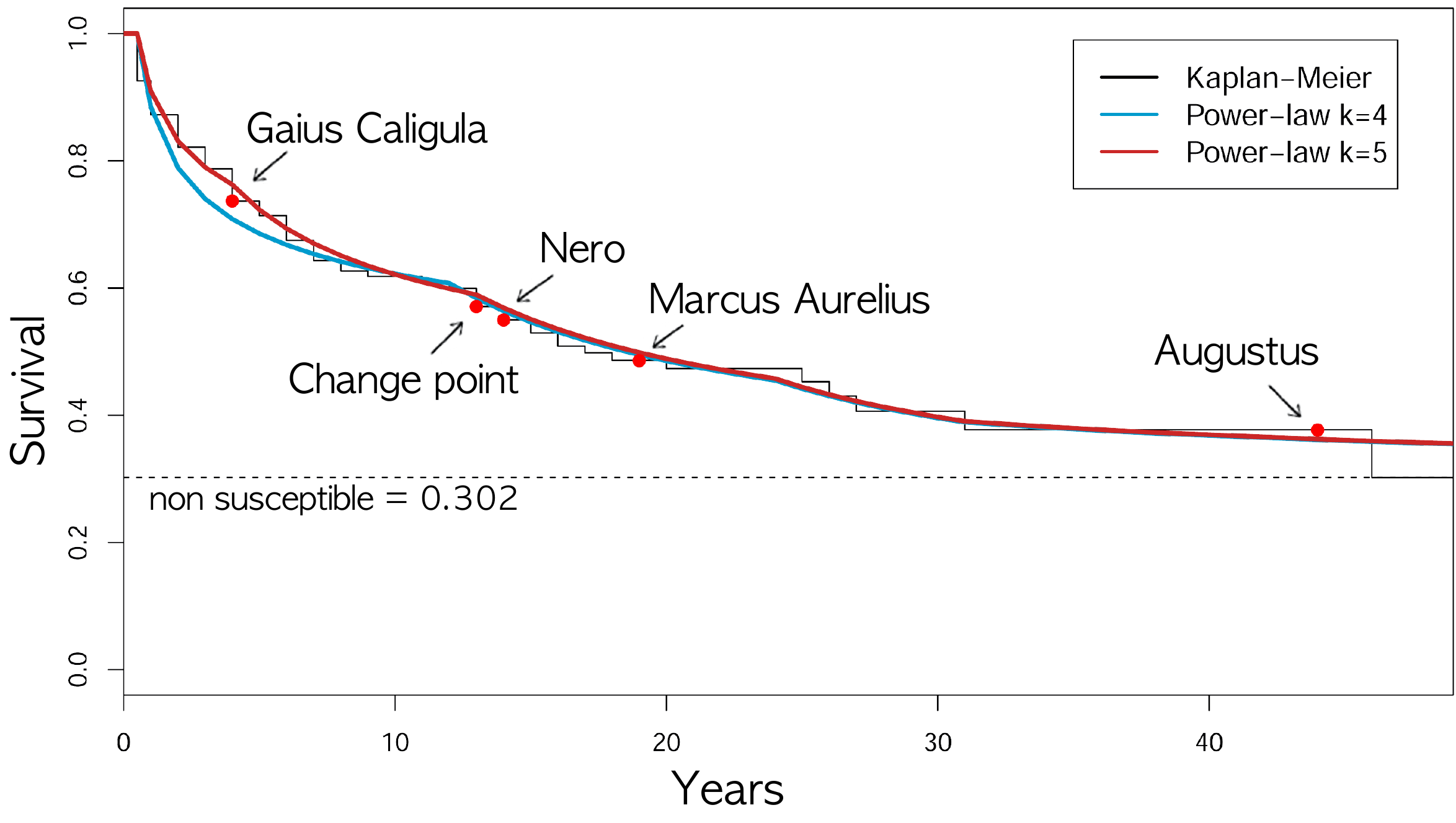}
	\caption{Power-law survivor (reliability) function of Roman emperors for all 175 Roman emperors and the nonparametric KM estimates. }\label{aplyalldyn}
\end{figure}

The details related to the generalized model with long-term survival and the inference procedure to estimate the parameters can be seen in section~\ref{Sec:methods}. This generalization is very flexible, and the fit improves following the number of change points. Figure \ref{aplyalldyn} presents the survival of the adjusted model using $k=4,5$ as well as the presence of long-term survival.
	
As shown in Figure \ref{aplyalldyn}, as the number of change points increases, we obtain a better fit for the proposed data. In fact, with $k=5$, the adjusted model almost overlaps the non-parametric function. In order to avoid overfitting, we consider the Akaike information criteria (AIC) to discriminate between the two models. In this case, the best fit among the chosen models returns the minimum value of AIC. While the model with $k=4$ returned an $AIC=594.10$, the model with $k=5$ yields $AIC=586.25$. Thus, this is the best model for describing the time-to-violent-death for all emperors.

\subsection{The time-to-violent-death of Western emperors}

\begin{figure}[!t]
	\centering
	\includegraphics[width=0.99\textwidth]{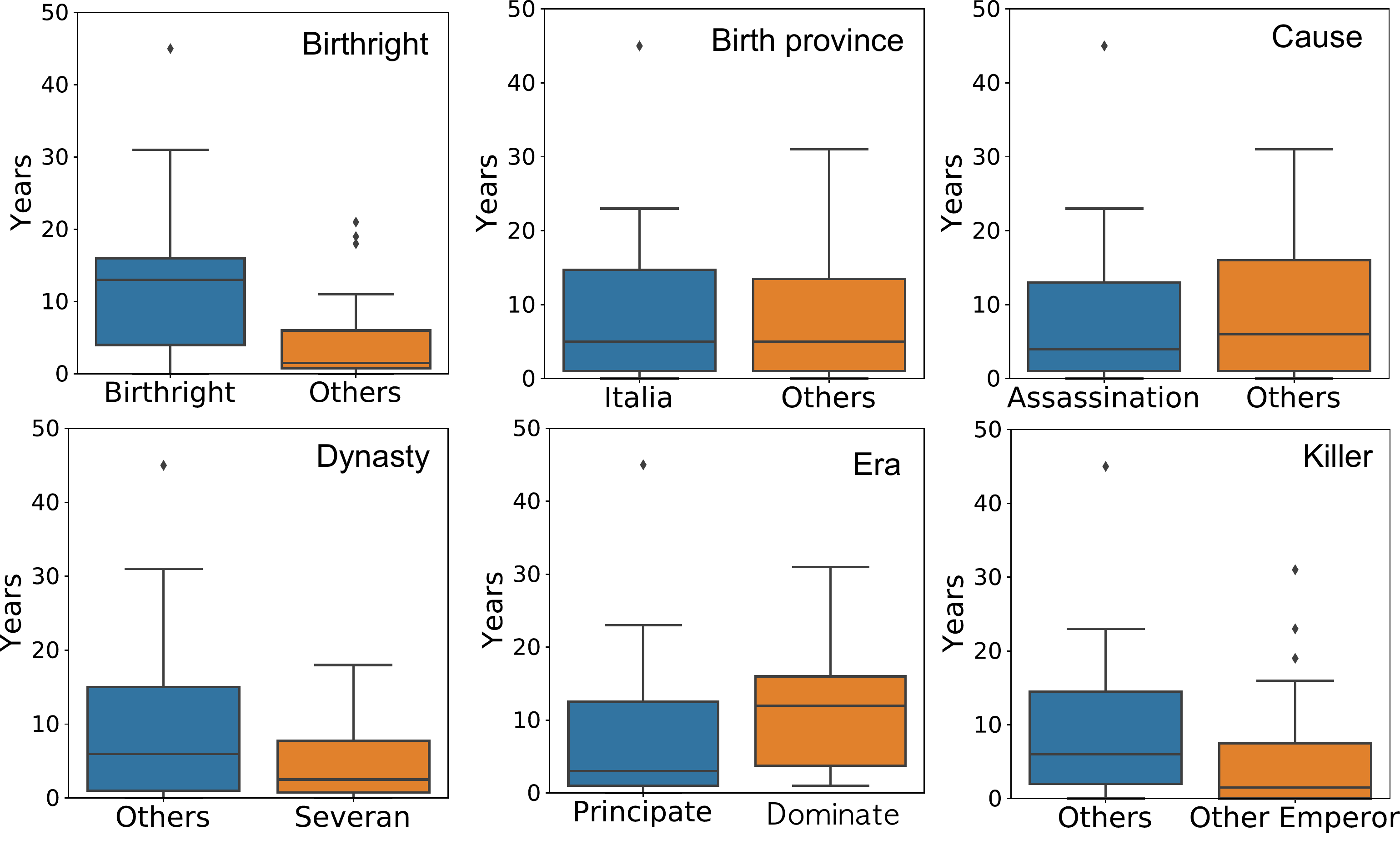}
	\caption{Box plot of emperors' attributes according to the time-to-death.}\label{Fig:attributes}
\end{figure}

We also analyze the influence of some emperors' attributes on the time-to-violent-death. These attributes are compiled from Wikipedia and previous works, including~\cite{Retief06, Cary75}, covering Western emperors from 26 BC to 395 AD. In the case of the method of accession to power (e.g., birthright, seized power, appointment by the senate, among others), we can see in figure~\ref{Fig:attributes} that emperors who came from birthright tend to rule longer than from other methods. To verify the significance of this difference, we consider a two-sided test for the null hypothesis that the average time-to-violent-death is independent of the method of accession to power. If the p-value is smaller than the threshold, .g. 1\% or 5\%, then we reject the null hypothesis of equal averages. From this hypothesis test,  we conclude that the difference between the time-to-violent-death of emperors that used different methods of accession to power is significant since the p-value $p<0.001$.  This result suggests that emperors who inherited the reign tend to have a more peaceful administration and an increased probability of having a natural death.
On the other hand, in terms of the birth province, there is no difference between the time-to-violent-death of emperors from Italy or other parts worldwide, like Syria or Spain ($p=0.54$). This finding indicates that Rome was very cosmopolitan, as the Empire was extensive, including many different cultures, traditions, and faiths. Thus, it is expected that the place of birth plays a small rule in the time-to-violent-death. 

The cause of death also does not influence the time-to-violent-death. Indeed, emperors, who died due to assassination, did not have shorter ruling periods ($p=0.8$), although almost 40\% of the emperors were assassinated. Related to this, emperors killed by other emperors also did not have a shorter reign period ($p=0.4$), as shown in the figure.

In terms of era, we verify the presence of a moderate difference ($p=0.08$) in terms of the reign period between the Principate. That is, the first period of the Roman Empire from the beginning of Augustus's reign in 27 BC, and Dominate, which started after the Crisis of the Third Century, in 284 AD.  The reigning period is more extended during the Dominate era, with an average time of ruling of 11 years, contrasting with seven years, verified in the Principate period covered in our data set. 

Regarding the dynasty, we also do not observe a significant difference between them in terms of the reign period ($p=0.2$). In figure~\ref{Fig:attributes}, we show the comparison between the Severan dynasty, which started with Septimius Severus, who rose to power as the victor of the Civil War of 193--197, and other dynasties. Although the Severan dynasty was disturbed by highly unstable family relationships and constant political turmoil~\cite{Cary75}, it did not significantly affect the time-to-violent-death. 

Most of the attributes considered in our analysis do not affect the time-to-violent-death, which suggests that other features, like those related to external factors like wars and political conflicts, should be considered to infer the causes of the short reign period of Roman emperors. This understanding is an exciting topic for further research.

\section{Discussion}

In this paper, we introduce a new survival model with $k$ change points and power-law distribution. The model is a modification of the traditional power-law model, in which we have included the adjustments to address the long-term survival and the new procedures to obtain the parameter estimates. The model has been developed to predict the reign period of Roman emperors, motivated by the analysis provided by Saleh~\cite{saleh2019statistical}.

In examining the time-to-violent-death of the emperors of the Roman Empire, we have observed that the risk is high when the emperor assumes the throne. This finding may be related to the struggles in handling the demands that the position requires and the lack of political skills of the new emperor. Our model suggests that the risk systematically decreases until 13 years of rule and then rapidly increases after this change point. The are many possible reasons for the change in the behavior of the risk, for instance,  their old enemies had regrouped, or new ones emerged \cite{saleh2019statistical}. After the change point, the risk decreases again.
Furthermore, the adjusted survival functions lead to $25\%$ of the emperor that would not die due to violent death. Therefore, being a Roman emperor was a hazardous occupation that led to 3-out-of-4 suffering from a cruel death during the Western Empire. Additionally, taking into consideration the Eastern Roman Empire, we observed that 2-out-of-3 emperors suffer from a cruel death.

The comparison with the model by Saleh \cite{saleh2019statistical} shows that the power-law model is more accurate to describe the variable of the interest, i.e., the time-to-violent-death of the emperors of the Roman Empire. This fact has important implications since the power-law distribution is ubiquitous in natural and artificial systems and is often related to critical events. Power-laws can be observed in earthquakes~\cite{bak2002unified, Mega03, kagan2010earthquake}, in solar radiophysics~\cite{Gary04}, stock price fluctuations \cite{gabaix2009power}, and tree-limb branching \cite{bentley2013empirical}. In the data set considered here, the lifetime of Roman emperors decay as a power-law,  which implies that short reigns are common, while high occurrences are rare. This behavior is similar to that observed in many systems, including earthquake, where huge earthquakes are rare. This result suggests that the time-to-violent-death does not have a typical scale, including the Roman Empire as an example of systems presenting scale invariances, such as the distribution of income, size of cities according to population, size of corporations, and word frequencies~\cite{clauset2009power}.

Our power-law survival model with $k$ change points is general and can be applied to model the hazard rate function is engineering, medicine, economy, and physics, especially in situations where we have a good indication that the tail of the distribution follows a power-law. However, the earlier occurrences may have a more slow decay. In addition to the proposed results drawn here, there are additional aspects of the new distribution that should be considered. For instance, the theoretical results assuming discrete data are currently under investigation. Under this scenario, the estimation procedures do not have closed-form expression and need to be further improved.  Thus, our results open several possibilities in applications and theoretical development, including models with cutoff and improvements by considering Bayesian methods.

\section{Methods}\label{Sec:methods}

This section contains most of the technical information related to the proposed model. We will discuss the mathematical properties such as the higher moments which allow us to obtain the mean, variance, among other measures. We also discuss the inferential procures to obtain the parameter estimates under the maximum likelihood estimators in the presence of complete and censored data.

\subsection{Mathematical Properties}\label{mathemaproppw}

The r-th moment is a useful function that can be used to obtain different measures related to the proposed distributions, including the mean, variance, and kurtosis. For the proposed distribution the r-th moment is given by

    \begin{equation}\label{eqmoc1}
		\begin{aligned}
    \mu_r&= \int_{x_{\min}}^{\infty}x^r f(x)dx  = \int_{x_{\min}}^{x_{c}}x^r\frac{\alpha_1-1}{x_{\min}}\left( \frac{x}{x_{\min}}\right)^{-\alpha_1}dx+\int_{x_{c}}^{\infty}C\frac{\alpha_2-1}{x_{c}}\left( \frac{x}{x_{c}}\right)^{-\alpha_2} dx \\& =  \frac{\left(x_{c}^{r-\alpha_1+1}-x_{\min}^{r-\alpha_1+1} \right)(\alpha_1-1)}{(r-\alpha_1+1)x_{\min}^{1-\alpha_1} }+\frac{(\alpha_2-1)x_{c}^{r-\alpha_1+1}}{(r-\alpha_2+1)x_{\min}^{1-\alpha_1} },
		\end{aligned}
    \end{equation}
where $r<\alpha-1$. 

Since the proposed distribution is an extension of the power-law distribution, when  $x_{c}\rightarrow\infty$ we have 
$$\lim_{x_{c}\rightarrow\infty}\frac{(\alpha_2-1)x_{c}^{r-\alpha_1+1}}{(r-\alpha_2+1)x_{\min}^{1-\alpha_1}}=0.$$

Additionally, applying the same limit in the left side of the  r-th moment we have
$$\lim_{x_{c}\rightarrow\infty}\frac{\left(x_{c}^{r-\alpha_1+1}-x_{\min}^{r-\alpha_1+1} \right)(\alpha_1-1)}{(r-\alpha_1+1)x_{\min}^{1-\alpha_1} }=\frac{(\alpha_1-1)}{(\alpha_1-r-1)}x_{\min}^{r}.$$

Hence, if $x_{c}\rightarrow\infty$ we have that $\mu_r=\frac{(\alpha_1-1)}{(\alpha_1-r-1)}x_{\min}^{r}$ which is the r-th moment of the standard power-law distribution.

The mean and variance are easily obtained from equation (\ref{eqmoc1}) by considering that $\mu=\mu_1$ and Var$(X)=\mu_1^2-\mu_2$. Additionally, measures such as skewness and kurtosis can also be obtained in closed-form using the higher orders of the obtained moments. 

The hazard function describe the instantaneous risk of failure at a
given of $x$ is given by
\begin{equation}
h(x)=\frac{\alpha_1-1}{x_{\min}}\left( \frac{x}{x_{\min}}\right)^{-1}\mathbb{I}[x_{\min},x_{c}]+\frac{\alpha_2-1}{x_{c}}\left( \frac{x}{x_{c}}\right)^{-1}\mathbb{I}(x_{c},\infty).
\end{equation}

The hazard function may exhibit different forms such as increasing, decreasing, unimodal, bathtub behavior among others. In this case, $\left( \frac{x}{x_{\min}}\right)^{-1}$ is an decreasing function multiplied by a constant term, hence, the hazard function initially has decreasing shape until $x_c$. Additionally, since $\left( \frac{x_c}{x_{\min}}\right)^{-1}<\left( \frac{x_c}{x_{c}}\right)^{-1}=1$, the risk increase rapidly and then decrease again as $\left( \frac{x}{x_{c}}\right)^{-1}$ is a decreasing function (see for instance Figure \ref{ajustkm}).

\subsection{Inference for the parameters}\label{infclasspw}

Usually, the parameters associated with the introduced distribution are unknown and need to be estimated. There are different estimation methods proposed in the literature. In this work, we have considered the maximum likelihood method. In this section, we discuss how to calculate the estimates assuming complete data. Further, we present the MLEs in the presence of random censoring, a general type of censoring that occurs in many situations.

\subsubsection{Maximum Likelihood Estimator}\label{comcesnmle}

The maximum likelihood method has been widely used due to its good asymptotic properties.
The estimates are obtained from the maximization of the likelihood function given by
\begin{equation*}
\begin{aligned}
L(\alpha_1,\alpha_2|x_{\min},x_{c},x)&=\prod_{i=1}^{n}\left(\frac{\alpha_1-1}{x_{\min}}\left( \frac{x_i}{x_{\min}}\right)^{-\alpha_1}\mathbb{I}[x_{\min},x_{c}]+C\frac{\alpha_2-1}{x_{c}}\left( \frac{x_i}{x_{c}}\right)^{-\alpha_2}\mathbb{I}(x_{c},\infty)\right)\\&=\prod_{i:x_i<x_c}\left(\frac{\alpha_1-1}{x_{\min}}\left( \frac{x}{x_{\min}}\right)^{-\alpha_1}\right)\prod_{i:x_i\geq x_c}\left(C\frac{\alpha_2-1}{x_{c}}\left( \frac{x}{x_{c}}\right)^{-\alpha_2}\right).
\end{aligned}
\end{equation*}
 
Instead of direct maximizing the likelihood function, it is straightforward to maximize the log-likehood function that is given by
\begin{equation*}
\begin{aligned}
l(\alpha_1,\alpha_2|x_{\min},x_{c},x)&=n_1\log(\alpha_1-1)-n_1(1-\alpha_1)\log(x_{\min})-\alpha_1\sum_{i:x_i<x_c}\log(x_i) +n_2\log(\alpha_2-1)\\& -n_2(\alpha_1-\alpha_2)\log(x_c)-n_2(1-\alpha_1)\log(x_{min})  -\alpha_2\sum_{i:x_i\geq x_c}\log(x_i).
\end{aligned}
\end{equation*}
From the expressions  $\frac{\partial}{\partial \alpha_1}l(\alpha_1,\alpha_2|x_{\min},x_{c},x)=0$, $\frac{\partial}{\partial \alpha_2}l(\alpha_1,\alpha_2|x_{\min},x_{c},x)=0$, the likelihood equations is given as
\begin{equation}\label{vere1}  
\hat\alpha_{1}=1+n_1\left(\sum_{i:x_i< x_c}\log\left(\frac{x_i}{x_{\min}}\right)+n_2\log\left(\frac{x_c}{x_{\min}}\right)\right)^{-1},
\end{equation}
\begin{equation}\label{vere2} 
\hat\alpha_{2}=1+n_2\left(\sum_{i:x_i\geq x_c}\log\left(\frac{x_i}{x_{c}}\right)\right)^{-1}.
\end{equation}

The maximum likelihood estimate is asymptotically normal distributed with a normal distribution given by
$ (\hat\alpha_1,\hat\alpha_2) \sim N((\alpha_1,\alpha_2),I^{-1}(\alpha_1,\alpha_2))) \mbox{ for } n \to \infty$, where $I^{-1}(\alpha_1,\alpha_2)$ is the Fisher information matrix with elements given by
\begin{equation}\label{var1a}
I^{-1}_{11}(\alpha_1,\alpha_2)=-E\left[\frac{\partial^2}{\partial \alpha^2} l(\alpha;\boldsymbol{t})\right]=\frac{(\alpha_1-1)^2}{n_1}, \quad
I^{-1}_{22}(\alpha_1,\alpha_2)=-E\left[\frac{\partial^2}{\partial \alpha^2} l(\alpha;\boldsymbol{t})\right]=\frac{(\alpha_2-1)^2}{n_2}
\end{equation}
and $I^{-1}_{12}(\alpha)=I^{-1}_{21}(\alpha)=0$.

\subsubsection{The presence of random censoring}\label{rancesnmle}

 In the presence of partial information the likelihood function need to be modified. Let $T_i$ be the lifetime of $i$th  emperor with censoring time $C_i$, which are assumed to be independent of
$T_i$s and its distribution does not depend on the parameters, the data set is represented by $\mathcal{D}=(x_i,\delta_i)$, where $x_i=\min(T_i,C_i)$ and $\delta_i=I(T_i\leq C_i)$. The likelihood function for $\boldsymbol\theta$ is given by
\begin{equation*}
\begin{aligned}
 L(\alpha_1,\alpha_2;x_{\min},x_{c},\boldsymbol{x,\delta})&=\prod_{i=1}^n f(x_i|\boldsymbol{\theta})^{\delta_i}S(x_i|\boldsymbol{\theta})^{1-\delta_i}
\\&=\prod_{i:x_i<x_c}\left(\frac{\alpha_1-1}{x_{\min}}\left( \frac{x}{x_{\min}}\right)^{1-\delta_i-\alpha_1}\right)\prod_{i:x_i\geq x_c}\left(C\frac{\alpha_2-1}{x_{c}}\left( \frac{x}{x_{c}}\right)^{1-\delta_i-\alpha_2}\right).
\end{aligned} 
\end{equation*}

The log-likehood function is given by
\begin{equation*}
\begin{aligned}
l(\alpha_1,\alpha_2|x_{\min},x_{c},x)&=d_1\log(\alpha_1-1)-d_1(1-\alpha_1)\log(x_{\min})+\sum_{i:x_i<x_c}(1-\delta_i-\alpha_1)\log(x_i) +d_2\log(\alpha_2-1)\\& +n_2(1-\alpha_1)\log\left(\frac{x_c}{x_{min}}\right)-d_2\log(x_c)  +\sum_{i:x_i\geq x_c}(1-\delta_i-\alpha_2)\log\left(\frac{x_i}{x_c}\right).
\end{aligned}
\end{equation*}
From the expressions  $\frac{\partial}{\partial \alpha_1}l(\alpha_1,\alpha_2|x_{\min},x_{c},x)=0$, $\frac{\partial}{\partial \alpha_2}l(\alpha_1,\alpha_2|x_{\min},x_{c},x)=0$, the likelihood equations are given as
\begin{equation}
\hat\alpha_{1}=1+d_1\left(\sum_{i:x_i< x_c}\log\left(\frac{x_i}{x_{\min}}\right)+n_2\log\left(\frac{x_c}{x_{\min}}\right)\right)^{-1},
\end{equation}
\begin{equation}
\hat\alpha_{2}=1+d_2\left(\sum_{i:x_i\geq x_c}\log\left(\frac{x_i}{x_{c}}\right)\right)^{-1}.
\end{equation}

Similar to the case of complete data the maximum likelihood estimate is asymptotically normal distributed with a normal distribution given by
$ (\hat\alpha_1,\hat\alpha_2) \sim N((\alpha_1,\alpha_2),I^{-1}(\alpha_1,\alpha_2))) \mbox{ for } n \to \infty$, where $I^{-1}(\alpha_1,\alpha_2)$ is the Fisher information matrix with elements given by
\begin{equation*}
 I^{-1}_{11}(\alpha_1,\alpha_2)=-E\left[\frac{\partial^2}{\partial \alpha^2} l(\alpha;\boldsymbol{t})\right]=\frac{(\alpha_1-1)^2}{d_1}, \quad
I^{-1}_{22}(\alpha_1,\alpha_2)=-E\left[\frac{\partial^2}{\partial \alpha^2} l(\alpha;\boldsymbol{t})\right]=\frac{(\alpha_2-1)^2}{d_2}
\end{equation*}
and $I^{-1}_{12}(\alpha)=I^{-1}_{21}(\alpha)=0$.

\subsection{Simulation analysis}

This section is devoted to comparing the quality of
our proposed approach in terms of minimum Bias and the mean-square error (MSE). The metrics are computed by
\begin{equation*}
\f{Bias}_{\alpha}=\frac{1}{N}\sum_{i=1}^{N}(\hat\alpha_{i}-\alpha) \ \ \mbox{ and } \ \ \f{MSE}_{\alpha}=\sqrt{\sum_{i=1}^{N}\frac{(\hat\alpha_{i}-\alpha)^2}{N}},
\end{equation*} 
where $N=200,000$ is the number of estimates obtained through the MLE.  An adequate estimation method will return the Bias and the MSEs closer to zero. The software R (R Core Development Team) was used to conduct this simulation study. The codes are available in the supplemental material.

\begin{figure}[!t]
	\centering
	\includegraphics[width=0.8\textwidth]{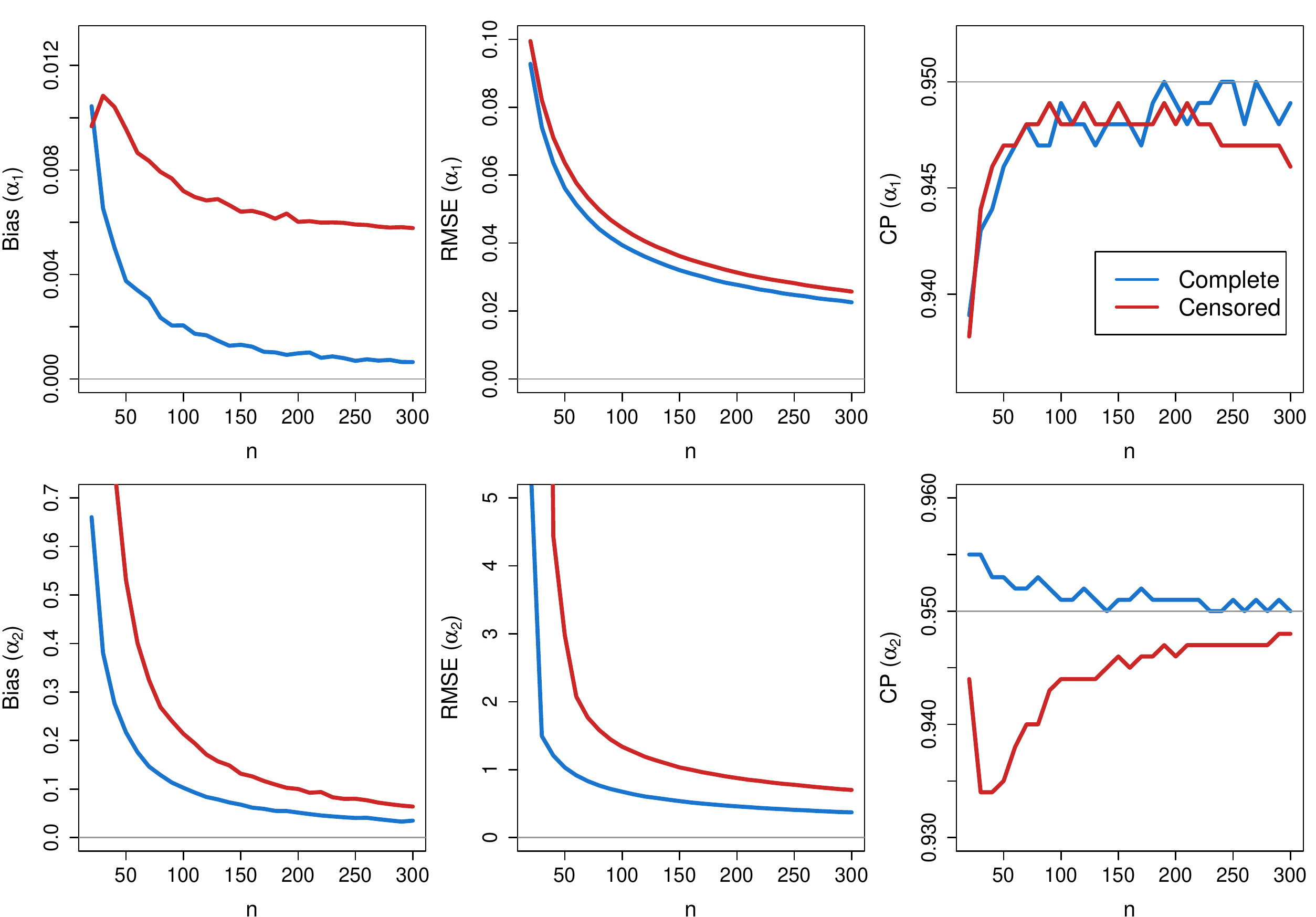}
	\caption{ Bias and RMSEs for $\alpha_1$ and $\alpha_2$ considering $x_m=0.5$, $x_c=13$, $\alpha_1=1.3$ and $\alpha_2=6$ for $N=200,000$ random samples and $n=(20,60,\ldots,300)$}\label{fsim1}
	\end{figure}

The pseudo-random samples from a power-law distribution with change phase can be easily generated by considering the quartile function that is given by
\begin{equation} 
x_u=\begin{cases}
x_m(1-u)^{1/(1-\alpha_1)} &  \text{if } u\geq(1-c) \\
x_c\left(\frac{1-u}{c}\right)^{1/(1-\alpha_2)} &  \text{if } u>(1-c),
\end{cases}
\end{equation}
where $u$ is generated from a uniform distribution with range (0,1). Hence, setting the parameters and repeating the process $n$ times we obtain a pseudo-random sample of size $n$ that is considered as complete data. To generate censored data an additional step is necessary. Since we have assumed that the censoring times $C_i$ are assumed to be independent of
$X_i$s and we sample a new value $y$ from a Uniform$(0,y_{\max})$, where $y_{\max}$ is the value that allows us to obtain the desirable proportion of censoring, then the observed time is the minimum $(x_u,y)$ and it is considered complete if the value is $x_u$ or censored if the value came from y. Here, we assume that the values are $x_m=0.5$, $x_c=13$, $\alpha_1=1.3$ and $\alpha_2=6$ and $y_{\max}$ is selected to obtain in average the proportion of censoring of $0.37$. The upper bound distribution Figure \ref{fsim1} displays the Bias and the RMSEs from the estimates of $\alpha$ obtained using the MC method. The horizontal lines in the figures correspond to Bias and RMSEs being one and zero respectively.

As shown in Figure \ref{fsim1}, the results related to $\alpha_1$ do not have large values, it can be observed that the bias is minimal for both complete and censored data, while $\alpha_2$ imposes more bias in the presence of small size samples. On the other hand, both Bias and RMSEs tend to zero when $n$ increases. Moreover, we obtain accurate coverage probability through
the CIs, since they cover the true values for the parameters even for small samples. Overall, these results show that the MLE estimators could be considered for estimating the parameters of the proposed model.

\subsection{A long-term survivor power-law distribution}\label{longtpwl}

Here we suppose that the population is split into two groups: those emperors that are not susceptible to the event of interest with probability $\pi$, and those who are susceptible (in risk) to the event with probability $(1-\pi)$. The event of interesting is ``a violent death''. The long-term survival function is then given by
\begin{equation}\label{fcp1}
S_{\rm pop}(x) = \pi+(1-\pi)S_0(x),    
\end{equation}
where $\pi\in(0,1)$ and $S_0(x)$ denotes the baseline survival function for the susceptible group in the population. 

This model is known as the standard long-term survival model and was introduced by \cite{berkson1952survival}. Assuming that the baseline survival function is given by our model, the obtained (unconditional) survival function \eqref{fcp1} is given by
\begin{equation}
S_{\rm pop}(x) = \pi+(1-\pi)\left( \frac{x}{x_{\min}}\right)^{1-\alpha_1}\mathbb{I}[x_{\min},x_{c}]+(1-\pi)C\left( \frac{x}{x_{c}}\right)^{1-\alpha_2}\mathbb{I}(x_{c},\infty).
\end{equation}

From the equation above we can obtain its related density function that is given by
\begin{equation}\label{fcp2}
f_{\rm pop}(x) = (1-\pi)\frac{\alpha_1-1}{x_{\min}}\left( \frac{x}{x_{\min}}\right)^{-\alpha_1}\mathbb{I}[x_{\min},x_{c}]+(1-\pi)C\frac{\alpha_2-1}{x_{c}}\left( \frac{x}{x_{c}}\right)^{-\alpha_2}\mathbb{I}(x_{c},\infty).
\end{equation}

Here, we have included an additional parameter of $\pi$ that needs to be estimated. Under the maximum likelihood estimators and assuming that the data have random censoring, the estimates are obtained from the maximization of the likelihood function given by

\begin{equation*}
\begin{aligned}
 L(\alpha_1,\alpha_2,\pi;x_{\min},x_{c},\boldsymbol{x,\delta})&=\prod_{i=1}^n f(x_i|\boldsymbol{\theta})^{\delta_i}S(x_i|\boldsymbol{\theta})^{1-\delta_i}\\&=\prod_{i:x_i<x_c}\left((1-\pi)\frac{\alpha_1-1}{x_{\min}}\left( \frac{x}{x_{\min}}\right)^{-\alpha_1}\right)^{\delta_i}\left(\pi+(1-\pi)\left( \frac{x}{x_{\min}}\right)^{1-\alpha_1}\right)^{1-\delta_i}\\
&\times\prod_{i:x_i\geq x_c}\left((1-\pi)C\frac{\alpha_2-1}{x_{c}}\left( \frac{x}{x_{c}}\right)^{-\alpha_2}\right)^{\delta_i}\left(\pi+(1-\pi)C\left( \frac{x}{x_{c}}\right)^{1-\alpha_2} \right)^{1-\delta_i}.
\end{aligned} 
\end{equation*}

The log-likelihood function that is used to find the estimates is given by
\begin{equation*}
\begin{aligned}
 l(\alpha_1,\alpha_2,\pi;x_{\min},x_{c},\boldsymbol{x,\delta})&=\sum_{i:x_i<x_c}\delta_i\log\left((1-\pi)\frac{\alpha_1-1}{x_{\min}}\left( \frac{x}{x_{\min}}\right)^{-\alpha_1}\right) +\\
 &+ \sum_{i:x_i<x_c}(1-\delta_i)\log\left(\pi+(1-\pi)\left( \frac{x}{x_{\min}}\right)^{1-\alpha_1}\right) +\\
 &+\sum_{i:x_i\geq x_c}\delta_i\log\left((1-\pi)C\frac{\alpha_2-1}{x_{c}}\left( \frac{x}{x_{c}}\right)^{-\alpha_2}\right)+\\
&+ \sum_{i:x_i<x_c}(1-\delta_i)\log\left(\pi+(1-\pi)C\left( \frac{x}{x_{c}}\right)^{1-\alpha_2} \right).
 \end{aligned} 
\end{equation*}

The maximum likelihood estimates are obtained  solving the expressions  $\frac{\partial}{\partial \alpha_1}l(\alpha_1,\alpha_2,\pi;x_{\min},x_{c},\boldsymbol{x,\delta})=0$, $\frac{\partial}{\partial \alpha_2}l(\alpha_1,\alpha_2,\pi$  $;x_{\min},x_{c},\boldsymbol{x,\delta})=0$ and  $\frac{\partial}{\partial \pi}l(\alpha_1,\alpha_2,\pi;x_{\min},x_{c},\boldsymbol{x,\delta})=0$. Unfortunately for this scenario the MLEs does not have closed-form expression and numerical methods such as Newton-Raphson need to be used to find the solution  of these
nonlinear equations.

\subsection{Long-term power-law model with k-1 change points}

Assuming that the baseline survival function is given by power-law distribution with $k-1$ change points, the related long-term survival model is given by
\begin{equation}
S_{\rm pop}(x) = \pi+(1-\pi)\sum_{i=1}^{k}\left( \frac{x}{x_{(i-1)}^{*}}\right)^{1-\alpha_i}C_{i-1}\mathbb{I}[x_{(i-1)}^{*},x_{(i)}^{*}].
\end{equation}

From the equation above we can obtain its related density function that is given by
\begin{equation}\label{fcp2l}
f_{\rm pop}(x) = (1-\pi)\sum_{i=1}^{k}\frac{\alpha_i-1}{x_{(i-1)}^{*}}\left( \frac{x}{x_{(i-1)}^{*}}\right)^{-\alpha_i}C_{i-1}\mathbb{I}[x_{(i-1)}^{*},x_{(i)}^{*}].
\end{equation}

Under the maximum likelihood estimators and assuming that the data have random censoring, the estimates are obtained from the maximization of the likelihood function given by

\begin{equation}
\begin{aligned}
 L(\boldsymbol{\Theta},x_{c},\boldsymbol{x,\delta})&=\prod_{i=1}^{n}\left((1-\pi)\sum_{j=1}^{k}\frac{\alpha_j-1}{x_{(j-1)}^{*}}\left( \frac{x_i}{x_{(j-1)}^{*}}\right)^{-\alpha_j}C_{j-1}\mathbb{I}[x_{(j-1)}^{*},x_{(j)}^{*}]\right)^{\delta_i}\left(\pi+(1-\pi)\sum_{i=1}^{k}\left( \frac{x_i}{x_{(j-1)}^{*}}\right)^{1-\alpha_j}C_{j-1}\mathbb{I}[x_{(j-1)}^{*},x_{(j)}^{*}]\right)^{1-\delta_i}\\&=\prod_{j=1}^{k}\left[\prod_{i:x_i\in[x_{(j-1)}^{*},x_{(j)}^{*}]}\left((1-\pi)C_{j-1}\left(\frac{\alpha_j-1}{x_{(j-1)}^{*}}\left( \frac{x_i}{x_{(j-1)}^{*}}\right)^{-\alpha_j}\right)\right)^{\delta_i}\left(\pi+(1-\pi)\left( \frac{x_i}{x_{(j-1)}^{*}}\right)^{1-\alpha_j}C_{j-1}\right)^{1-\delta_i}\right].
\end{aligned} 
\end{equation}

The log-likelihood function that is used to find the estimates is given by
\begin{equation*}
\begin{aligned}
 l(\boldsymbol{\Theta},x_{c},\boldsymbol{x,\delta})=&\sum_{j=1}^{k}\left[\sum_{i:x_i\in[x_{(j-1)}^{*},x_{(j)}^{*}]}\delta_i\log\left((1-\pi)C_{j-1}\left(\frac{\alpha_j-1}{x_{(j-1)}^{*}}\left( \frac{x_i}{x_{(j-1)}^{*}}\right)^{-\alpha_j}\right)\right)\right] \\& +\sum_{j=1}^{k}\left[\sum_{i:x_i\in[x_{(j-1)}^{*},x_{(j)}^{*}]}(1-\delta_i)\log\left(\pi+(1-\pi)\left( \frac{x_i}{x_{(j-1)}^{*}}\right)^{1-\alpha_j}C_{j-1}\right)\right].
\end{aligned} 
\end{equation*}

\section{Acknowledgements}
Pedro L. Ramos acknowledges support from the S\~ao Paulo State Research Foundation (FAPESP Proc. 2017/25971-0). Luciano da F. Costa thanks CNPq (grant no.  307085/2018-0) for sponsorship.  This work has benefited from FAPESP  grant 15/22308-2. Francisco Rodrigues acknowledges financial support from CNPq (grant number 309266/2019−0). Francisco Louzada acknowledges support from the S\~ao Paulo State Research Foundation (FAPESP Processes 2013/07375-0) and CNPq (grant no.  301976/2017-1).

\bibliographystyle{plain}
\bibliography{references}

\end{document}